\documentclass[11pt,a4paper]{article}

\usepackage[T1]{fontenc}
\usepackage[utf8]{inputenc} 
\usepackage{graphicx}
\usepackage{amsmath,amssymb}
\usepackage{cite}
\usepackage[colorlinks=true,linkcolor=blue,citecolor=blue,urlcolor=blue]{hyperref}
\usepackage{orcidlink}
\usepackage{QCDEX2026}

\begin{document}

\noindent\includegraphics[width=0.4\textwidth]{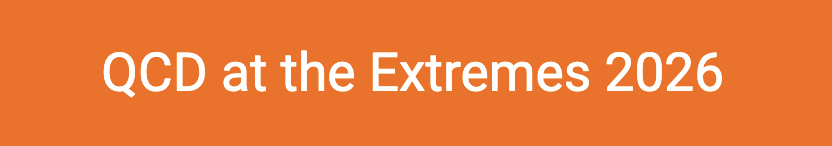}\par
\vspace{0.6em}
\begin{center}
  {\LARGE\bfseries New unintegrated gluon density in proton and nuclei tested with small $x$ processes at modern colliders\par}
  \vspace{0.8em}
  {\large
    M.A. Malyshev$^{1,*}$\orcidlink{0000-0002-8274-0277}
    ,\quad A.V. Lipatov$^{1}$
    \quad and\quad G.I. Lykasov$^{2}$\orcidlink{0000-0002-1544-6959}\par}
  \vspace{0.5em}
  {\small
    $^{1}$Skobeltsyn Institute of Nuclear Physics, Lomonosov Moscow State University, 119991 Moscow, Russia\par
    $^{2}$Joint Institute for Nuclear Research, 141980 Dubna, Moscow region, Russia\par}
  \vspace{0.35em}
  {\small $^{*}$Corresponding author: \href{mailto:malyshev@theory.sinp.msu.ru}{malyshev@theory.sinp.msu.ru}\par}
  \vspace{0.35em}
  {\small\itshape refereed by the conveners of QCDEX2026\par}
\end{center}
\vspace{0.5em}



\begin{abstract}
We present a new unintegrated, or transverse momentum dependent (TMD) gluon density in proton and nuclei. Parameters of the new function have been obtained from a fit on inclusive soft charged hadron production and structure functions at low scales, and from comparison with data on a number of processes at higher scales results taken at LHC and HERA, such as heavy quarks production, prompt photon photoproduction and Higgs boson production. A modification employing the geometric scaling approach has been made to extend the obtained parton density to the case of light and heavy nuclei. As a test of the nuclear TMD distribution we perform a calculation of the charged hadrons multiplicity spectra in pPb collisions at the LHC energies.
\end{abstract}


\vspace{1.0cm}

\newpage

Calculation of cross sections of particle production in hadron collisions relies on factorization theorems which allow to disentangle the non-perturbative physics of large distances and the parton scattering, which can be calculated order by order within the perturbation theory as the running strong coupling $\alpha_s(\mu_R^2)$ is small enough. The non-perturbative part is encoded in the parton distribution functions (PDFs), which are solutions of equations taking into account logarithmically enhanced terms arising during parton emissions in the course of the evolution on the parton subprocess scale $\mu^2$. If one considers only terms containing $\ln^n(\mu^2/\Lambda_\text{QCD}^2)$, one obtains the Dokshitzer-Gribov-Lipatov-Altarelli-Parisi (DGLAP) equations~\cite{DGLAP:1,DGLAP:2,DGLAP:3,DGLAP:4}, solutions of which are collinear PDFs $f(x,\mu^2)$. Such functions can then be convoluted with parton-level hard-scattering cross sections to calculate the cross section of a process. This procedure leads to the collinear factorization approach~\cite{Collins:2011}. 

At high energies terms enhanced by logarithms $\ln^n(\mu^2/s)$, with $s$ being the hadron collision center-of-mass energy squared, become important. The summation of such contributions can be implemented in non-collinear evolution equations, such as Balitsky-Fadin-Kuraev-Lipatov (BFKL)~\cite{BFKL:1,BFKL:2,BFKL:3} or Catani-Ciafaloni-Fiorani-Marchesini (CCFM)~\cite{CCFM:1,CCFM:2,CCFM:3,CCFM:4} equations. As solutions of such equations one gets PDFs, which start to depend on the transverse momentum of a parton, $f(x,\mathbf k_T^2,\mu^2)$. These functions are called unintegrated, or transverse momentum dependent (TMD) PDFs. Frequently used unintegrated PDFs are Kimber-Martin-Ryskin~\cite{KMR-LO:1,KMR-LO:2,KMR-NLO}, Jung-Hautmann 2013~\cite{JH2013} and Parton Branching~\cite{PB} TMDs. A framework to use the non-collinear PDFs is the high energy, or $k_T$-factorization approach (for a review see~\cite{TMD-review-our}). An advantage of this method is that it allows to take into account a number of higher orders perturbative corrections with the unintegrated PDFs, with no need to calculate higher order diagrams. 

The TMD PDF are rather badly studied. For example, a popular library to access collinear PDFs, \textsc{lhapdf}~\cite{lhapdf}, contains more than 1500 PDFs, while its analog for TMD PDFs, \textsc{tmdlib}~\cite{TMDLib2}, has less than 100 different TMDs. Concerning nuclear TMD PDFs, which can be used to describe also nuclei collisions, the situation is much worse, because very few examples exist~\cite{PB-nucl,LMKC,LK-nucl,MRW-nucl}. In this work we expand a recently introduced TMD PDF in proton LLM-2024~\cite{LLM-2024, LLM-QFTHEP25} to nuclei and test it with the process of charged hadron production at small transverse momenta, which has been studied at LHC in the recent years.

Let us briefly recall our procedure to introduce the unintegrated parton density. We start with an initial distribution at a small scale and then we expand it to large $\mu^2$ using the CCFM evolution equation. The initial distribution for the gluon TMD at low $\mu^2$ and small $x$ was chosen in a way to describe data on soft hadrons ($\pi$ and $K$-mesons) multiplicity spectra within the modified quark-gluon string model~\cite{LLM-2022}:
\begin{gather}
  f_{g}(x, {\mathbf k}_T^2) = c_g (1-x)^{b_g} \sum_{n = 1}^3 c_n \left[R_0(x)
    |{\mathbf k}_{T}|\right]^n \exp \left(-R_0(x)
  |{\mathbf k}_{T}|\right), \nonumber \\ 
  b_g = b_g(0) + {4C_A\over \beta_0}\ln{\alpha_s(Q_0^2)\over
    \alpha_s({\mathbf k}_{T}^2)},
  \label{eq:OurGluon}
\end{gather}
\noindent
where $R_0^2=1/Q_0^2\,(x/x_0)^\lambda$, $C_A = N_C$, $\beta_0 = 11 - 2N_f/3$, $Q_0=2.2$~GeV, and $Q_s\equiv 1/R_0$ can be taken as the saturation scale (see below). The phenomenological parameters
 $\lambda$, $x_0$, $c_n$, $c_g$ and $b_g(0)$ were determined from a fit on various LHC and HERA data (such as Higgs boson, prompt photon, heavy quarks production, etc.)~\cite{LLM-small-x-fit,LLM-2024,LLM-QFTHEP25}. 
 
A number of the parameters were extracted from a fit on the charged hadron multiplicities measured at LHC at low transverse momenta. At low scales perturbative calculations are unreliable, so we use the modified quark-gluon string model (mQGSM), developed in~\cite{mQGSM-1,mQGSM-2}. In this model the colliding protons are considered at low scales as systems consisting of three
valence quarks and gluon medium with the wave function 
$|\Psi_g|^2\sim f_g^{(0)}(x, {\mathbf k}_T^2, \mu_0^2)$.    
The spectrum of a hadron $h$ production $\rho(x,p_T)$ can be written as: 
\begin{gather}
 \rho(x,p_T)~=~\rho_q(x,p_T)~+~\rho_g(x,p_T),
\label{def:rho_qg}
\end{gather}
\noindent
where
\begin{gather}
\rho_q(x,p_T)=|f^{(0)}_{3q}|^2\otimes D_{q/qq\rightarrow h}\times\int d^2k_T dz f_g^{(0)}(z, {\mathbf k}_T^2, \mu_0^2),
\label{def:rho_q}
\end{gather}
with $D_{q/qq\rightarrow h}$ being the fragmentation function of quarks or diquarks to the hadrons.
The second term, $\rho_g(x,p_T)$, can be presented as convolution of 
$f_g^{(0)}$ with the gluon fragmentation function (FF), $D_{g\rightarrow h}$, and the integral of $|f^{(0)}_{3q}|^2$ over the 
phase spase, can be approximately taken as the inelastic $pp$ cross-section $\sigma_\text{in}^{pp}$:
\begin{gather}
\rho_g(x,p_T)=f^{(0)}_{g}\otimes D_{g\rightarrow h}\times\sigma_\text{in}^{pp}.
\label{def:rho_g}
\end{gather}

The hadron spectrum at low $x$ and $p_t$ can then be written as:

\begin{gather}
 \rho(x,p_T)\equiv E {d^3\sigma \over d^3 {\bf p}} \equiv {1\over \pi} {d^3 \sigma \over d^2 p_T dy} = \sigma_1 \phi_q(s,x,p_T) + \sigma_\text{in} \phi_g(s,x,p_T),
  \label{eq:HadronSpectrum}
\end{gather}
\noindent
where $E$, ${\bf p}$ and $y$ are the energy, $3$-dimensional momentum and rapidity of the produced hadron
having the transverse momentum $p_T \equiv |{\mathbf p}_T|$, $\sigma_1$ is the cross section of the one-Pomeron exchange and 
$\sigma_\text{in}$ is the inelastic $pp$ cross section. 

The FFs can be taken as
$D_{a\rightarrow h}(z, |{\mathbf {\tilde p}}_T|) = G_{a\rightarrow h}(z) I_h^a(|{\mathbf {\tilde p}}_T|)$, where $a$ stands for gluon, quark or diquark, and $I_h^a(|{\mathbf {\tilde p}}_T|)$
reads:
\begin{gather}
  I_h^a(|{\mathbf {\tilde p}}_T|)=\frac{B_h^a}{2\pi}\exp(-B_h^a
  |{\mathbf {\tilde p}}_T|).
  \label{def:FF_T}
\end{gather}
\noindent
Here $\tilde p_T = p_T - z k_T$ with $p_T$ and $k_T$
being the transverse momenta of the produced hadron $h$ and the parton, respectively. The collinear FFs $G_{a\rightarrow h}(z)$ were taken from~\cite{QGSM-FFs}. The slope parameters $B_h^a$ can be also extracted from a fit on the data. For proton collisions the slope parameters were taken the same for pions and kaons in~\cite{LLM-2022}, since the data did not allow a distinctive analysis for $\pi$ and $K$-mesons.

\begin{figure}[htbp]
  \centering
  \includegraphics[width=15cm]{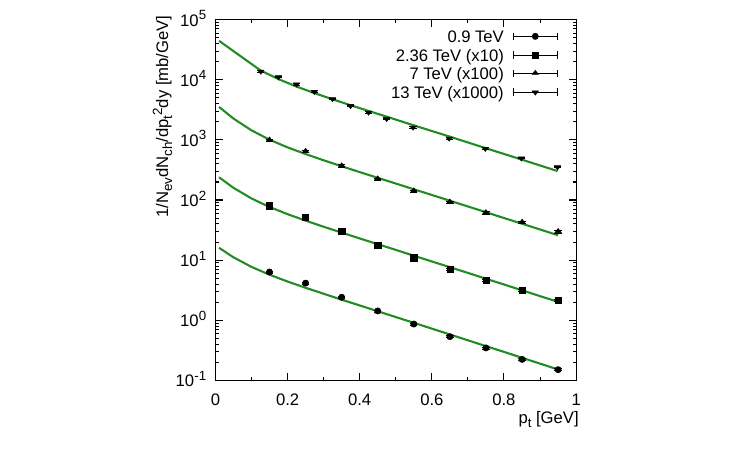}
  \caption{Transverse momentum distributions of soft charged
      hadrons produced in $pp$ collisions at different LHC energies
      in the mid-rapidity region. The solid green line corresponds to the results obtained with LLM TMD using mQGSM. The experimental data are from
      \cite{SoftHadronData:1,SoftHadronData:2,SoftHadronData:3}.}
\label{softhadrons}
\end{figure}

\begin{figure}[htbp]
  \centering
	  \includegraphics[width=5.0cm]{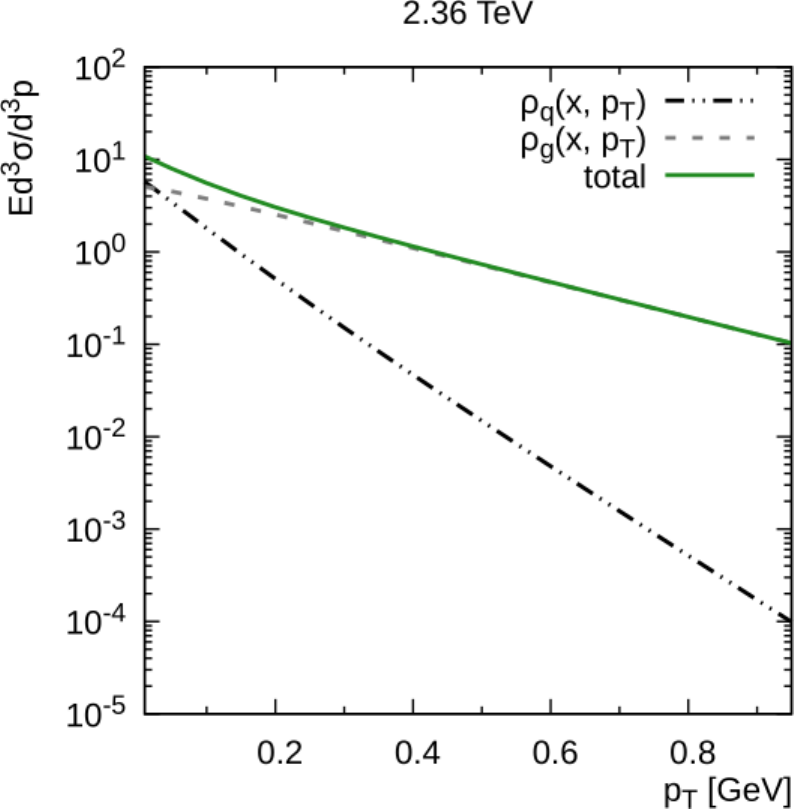}
	  \includegraphics[width=5.0cm]{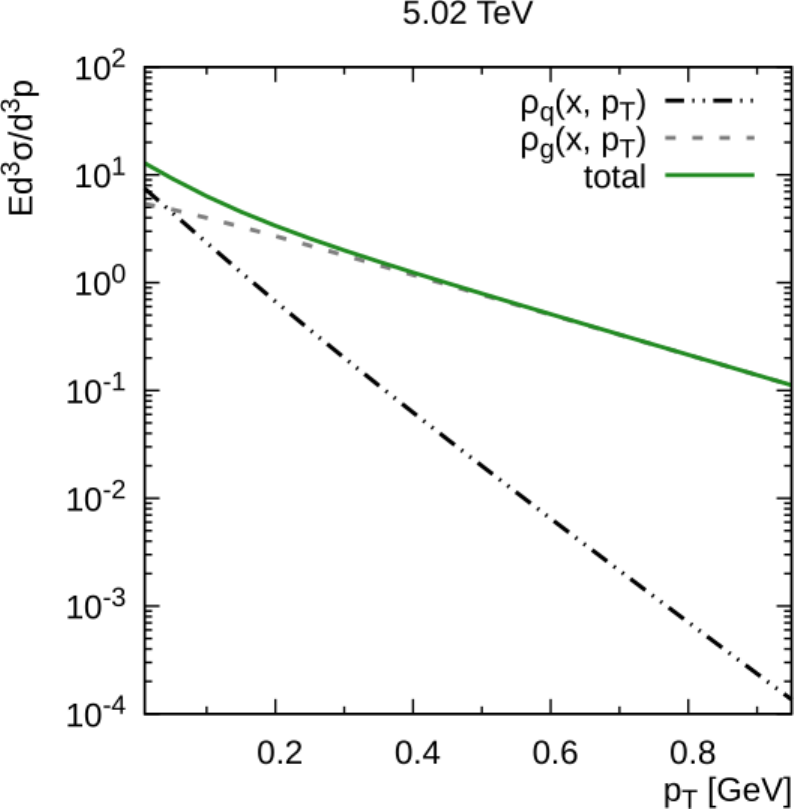}
	  \includegraphics[width=5.0cm]{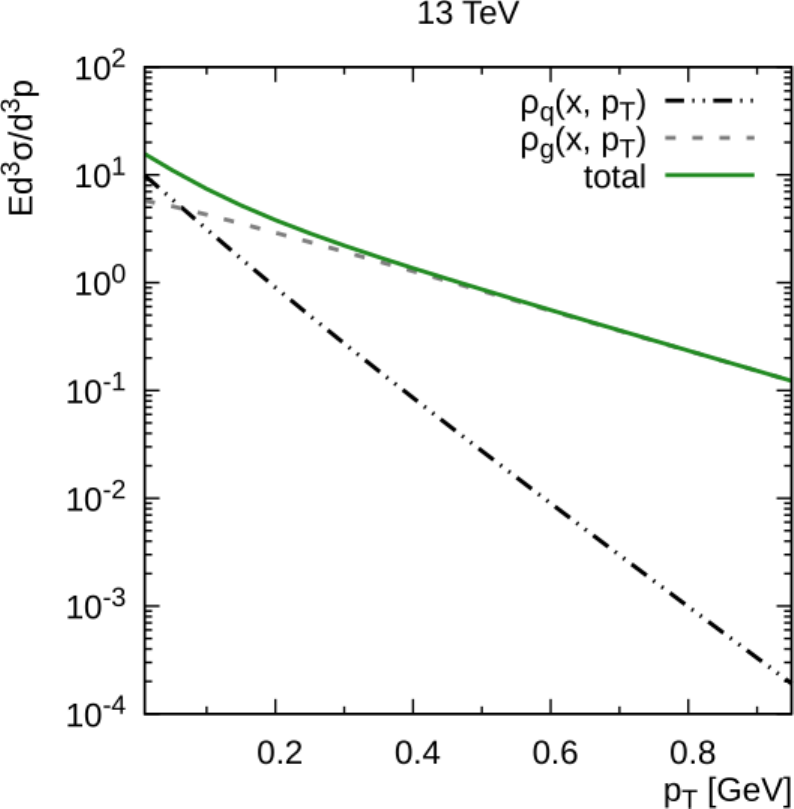}
	  \caption{Quark/diquark (black dash-dotted) and gluon (gray dashed)
          contributions to the inclusive spectrum of soft charged hadrons 
          calculated at $\sqrt{s} = 2.36$, $5.02$ and $13$~TeV. Sum of all contributions is represented by the green line.}
		\label{fig1}
\end{figure}

The results of the fit are shown in Fig.~\ref{softhadrons}. In Fig.~\ref{fig1} we show also separately quark and gluon contributions to the cross section at different energies. One can see that the quark contribution is relevant only at very low $p_T$.

\begin{figure}[htbp]
  \centering
\includegraphics[width=15cm]{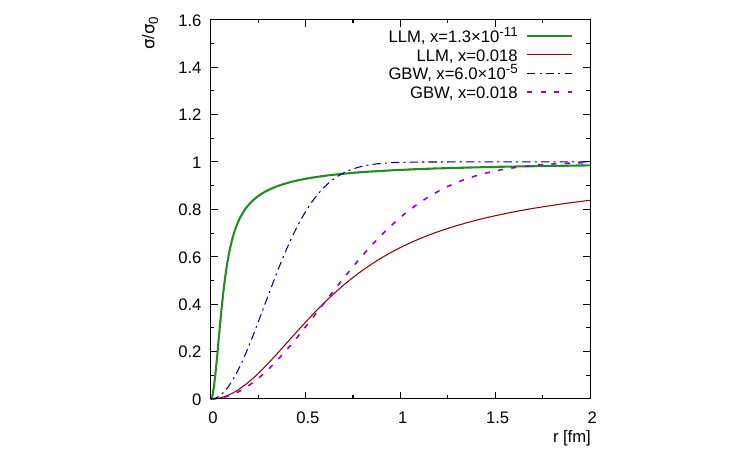}
\caption{The effective dipole cross section $\hat \sigma(x, r^2)$
    normalized by $\sigma_0$ 
  calculated as a function of the dipole size $r$ at different values of $x$.
}
\label{saturation}
\end{figure}

The LLM TMD gluon distribution posesses the saturation property, similarly to the Golec-Biernat--Wusthoff (GBW) distribution~\cite{GBW-1,GBW-2}. So, the dipole cross section can be calculated using the two gluon exchange approximation between the color dipole $q{\bar q}$
  and the proton debris:
\begin{gather}
  \hat{\sigma}(x,r^2) = \frac{4\pi^2 \alpha_s}{3}\int\frac{d{\mathbf k}_T^2}
      {{\mathbf k}_T^2}\left\{1-J_0(|{\mathbf k}_T| r)\right\}
      f_g(x, {\mathbf k}_T^2),
  \label{def:GBW_dpcrsect}
\end{gather}
\noindent
where $r$ is the dipole size, $J_0$ is the Bessel function of zeroth order and
  $\alpha_s$ is taken fixed. The result is shown in Fig.~\ref{saturation}. For comparison we show also the results obtained with the GBW TMD. We observe a saturation dynamics for LLM-2024 at large $r$, which is $x$-dependent. 

Our next step was to expand the TMD to nuclei.
Two main approaches for obtaining nuclear PDF are used: 1) analogously to the proton case, one chooses an initial distribution, which is evolved accorging to (modified) evolution equations, and the parameters of the distribution are fitted to existing data~\cite{EPPS21,nNNPDF3,nCTEQ,nIMP}; 
2) a nuclear PDF is obtained from a proton one using a special modification scheme (Kulagin-Petti model~\cite{KulaginPetti-1,KulaginPetti-2}, rescaling model~\cite{RescalingModel-1:1,RescalingModel-1:2,RescalingModel-1:3,RescalingModel-2:1,RescalingModel-2:2} etc.).
In the present work we choose the latter way and obtain a nuclear PDF form the proton LLM-2024 using the concept of the geometric scaling~\cite{GeometryScaling-1,GeometryScaling-2}. We find it useful in case of $k_T$-factorization, because it is valid in the low $x$ region, where the nuclear shadowing plays the dominant role, and the geometric scaling approach is specially good at describing this effect (see, for instance,~\cite{GeometryScaling-1,nGBW}).

The idea of the geometric scaling comes from data on the virtual photon scattering on the proton cross section,
$\sigma^{\gamma^* p}(x,Q^2)$, which
can be described in a wide region of $Q^2$
with a single scaling variable $\tau = Q^2/Q_s^2(x)$.
For nuclear cross sections $\sigma^{\gamma^* A}(x,Q^2)$ one can observe similar behaviour~\cite{GeometryScaling-2}.
It can be demonstrated that
\begin{gather}
  {\sigma^{\gamma^* A}(\tau) \over \pi R_A^2} = {\sigma^{\gamma^* p}(\tau) \over \pi R_p^2},
\label{eq-GeometryScalingA}
\end{gather}
\noindent
where $R_h$ is the radius of the hadronic target and $\tau = \tau_{A,p} \equiv Q^2/Q_{s,\,A,p}^2$.
The $A$-dependence of $Q_{s,\,A}(x)$, can be taken in the following form~\cite{GeometryScaling-2}:
\begin{gather}
  Q^2_{s,\,A}(x) = Q^2_s(x) \left( {A \pi R_p^2 \over \pi R_A^2} \right)^{1\over \delta}
\label{eq-SaturationScaleA}
\end{gather}
\noindent
where $\delta$ and $\pi R_p^2$ are free parameters to be fitted from data.
So, the substitution $Q_s^2(x) \to Q_{s, \, A}^2(x)$~\cite{GeometryScaling-2}
and applying the overall normalization factor $R_A^2/R_p^2$ transform the proton gluon density to a nuclear one. The proton LLM distribution shows the saturation behaviour with the $x$-dependent saturation scale $Q_s\equiv 1/R_0$, so it can be used to obtain a nuclear TMD with this method.

\begin{figure}[htbp]
  \centering
\includegraphics[width=5.0cm]{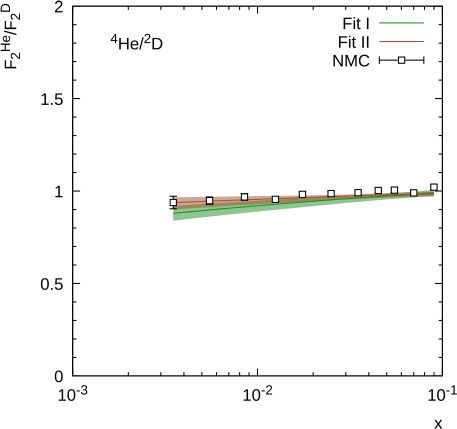}
\includegraphics[width=5.0cm]{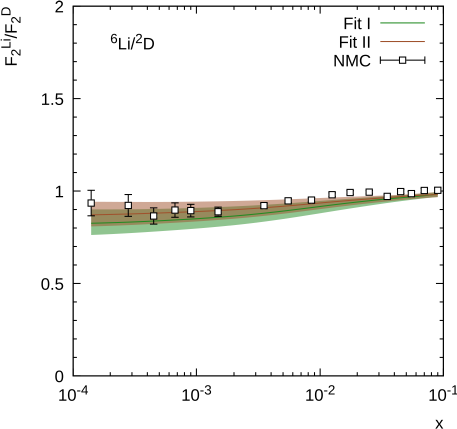}
\includegraphics[width=5.0cm]{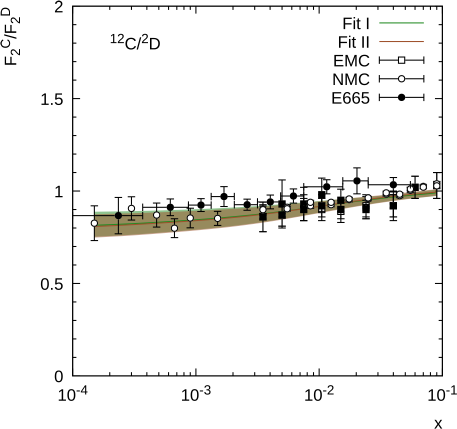} 
\includegraphics[width=5.0cm]{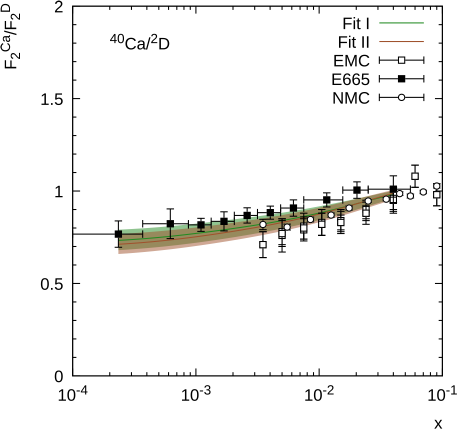}
\includegraphics[width=5.0cm]{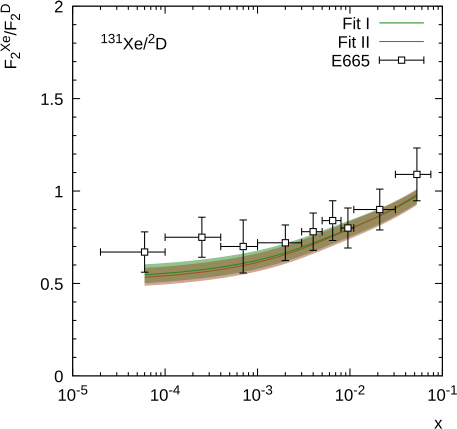}
\includegraphics[width=5.0cm]{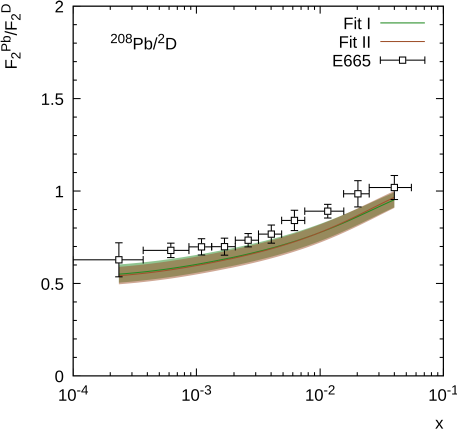}
\caption{The fit results of structure function ratios
$F_2^{A}(x, Q^2)/F_2^{D}(x, Q^2)$ between different nuclei target $A$ and deuterium. 
Shaded bands represent the estimated uncertainties of the fit procedure.
Experimental data are from
NMC, EMC, and E665~\cite{F2A:1,F2A:2,F2A:3,F2A:4,F2A:5,F2A:6,F2A:7}.}
\label{F2AF2D} 
\end{figure}

A fit on $F_2^A(x,Q^2)/F_2^D(x,Q^2)$ data can give us the scaling parameters $\pi R_p^2$ and $\delta$~\cite{nLLM-2024}. For the nuclear radius we take two parametrizations known from the literature~\cite{GeometryScaling-2, RA-Textbook}:
\begin{equation}
	R_A = \left(1.12 A^{1/3} - 0.86 A^{-1/3}\right)\text{\,fm},
	\label{RA}
\end{equation}
\noindent 
and 
\begin{equation}
	R_A = \left( 1.12 A^{1/3} - 0.5 \right)\text{\,fm}.
	\label{RA-Kapitonov}
\end{equation}
The fit obtained with (\ref{RA}) is denoted as 'Fit I' 
and with (\ref{RA-Kapitonov}) as 'Fit II'.
Our fits for $\pi R_p^2$ and $\delta$ are
$\pi R_p^2 = 1.74 \pm 0.20 $~fm$^2$, $\delta = 0.751 \pm 0.026$ with $\chi^2/n.d.f.= 2.28$ for Fit I 
and $\pi R_p^2 = 1.86 \pm 0.20$~fm$^2$, $\delta = 0.740^{+0.041}_{-0.021}$ with $\chi^2/n.d.f. = 2.19$ for Fit II. Some of our results are shown in Fig.~\ref{F2AF2D}.
The both fits show a good description of the data.

Having obtained the nuclear TMD PDF, we test it with the charged hadron multiplicities spectra measured at low $p_T$ in proton-lead collisions~\cite{nLLM-test}. To do it we have to only slightly change the mQGSM scheme described earlier. We use the same formula~(\ref{eq:HadronSpectrum}), but the gluon distribution in~(\ref{def:rho_g}) is now replaced by the nuclear one for one of the colliding particles. As we showed in Fig.~\ref{fig1}, quark contribution is small in comparison with the gluon one, so we keep it the same as in the proton case, so the nuclear modification comes only to the gluon part. Also the fragmentation function parameters, $B^a_h$ in~\ref{def:FF_T}, can also change in the nuclear collisions, so they can be modified to better describe data. Moreover, the data~\cite{ChargedHadron5-CMS,ChargedHadron5-ALICE} allow to distinguish between pion and kaon production and determine their fragmentation parameters separately.

\begin{figure}[htbp]
  \centering
        \includegraphics[width=5.0cm]{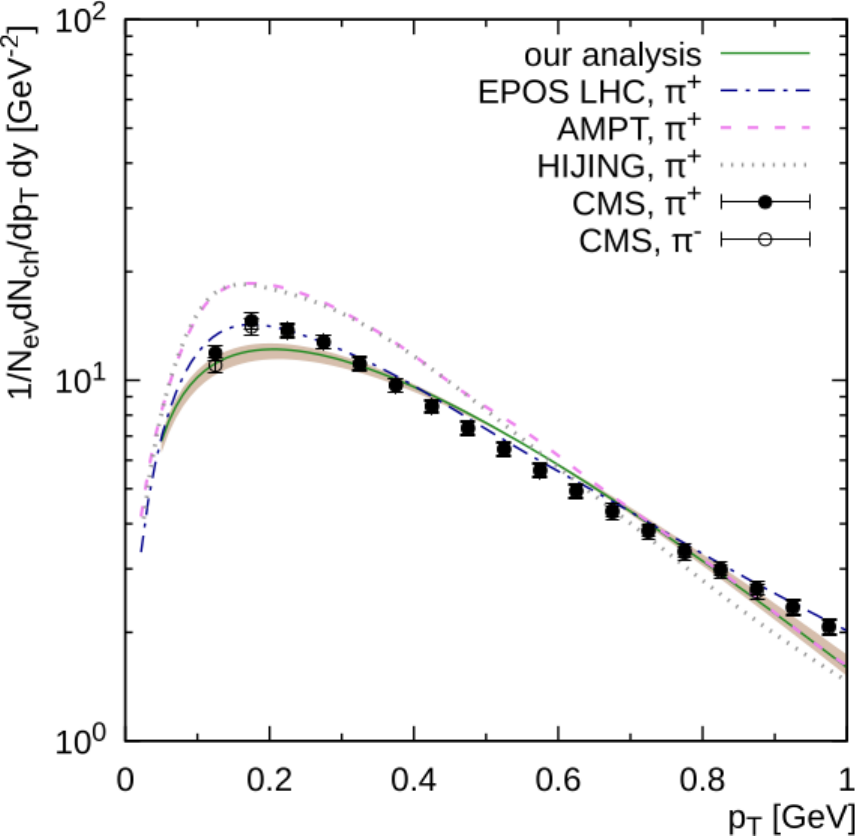}
        \includegraphics[width=5.0cm]{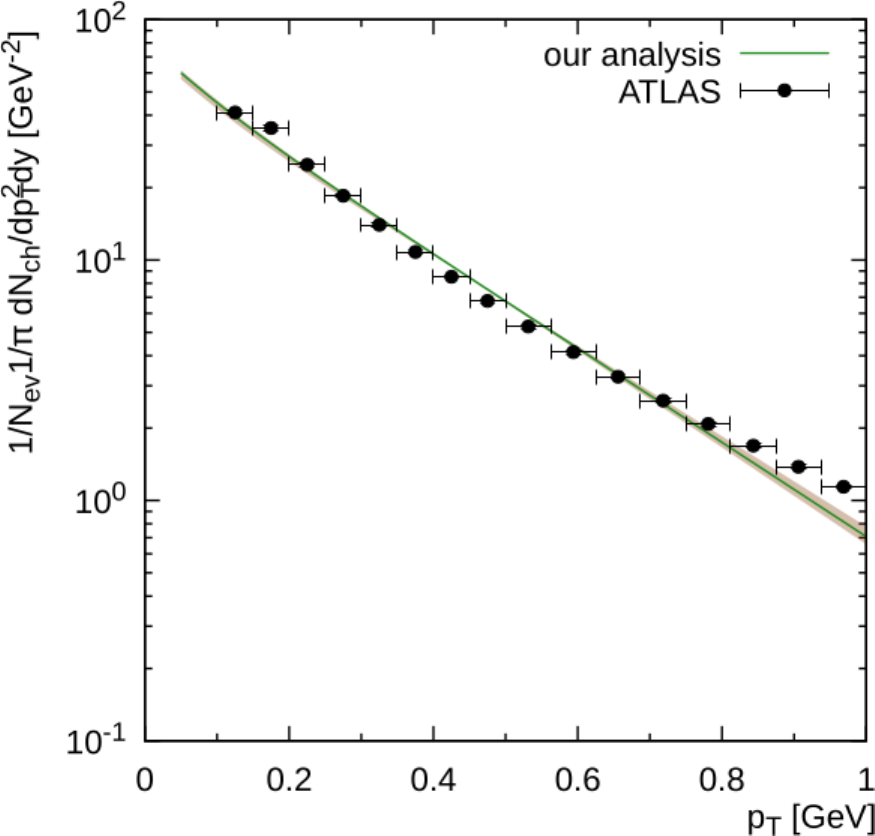}
        \includegraphics[width=5.0cm]{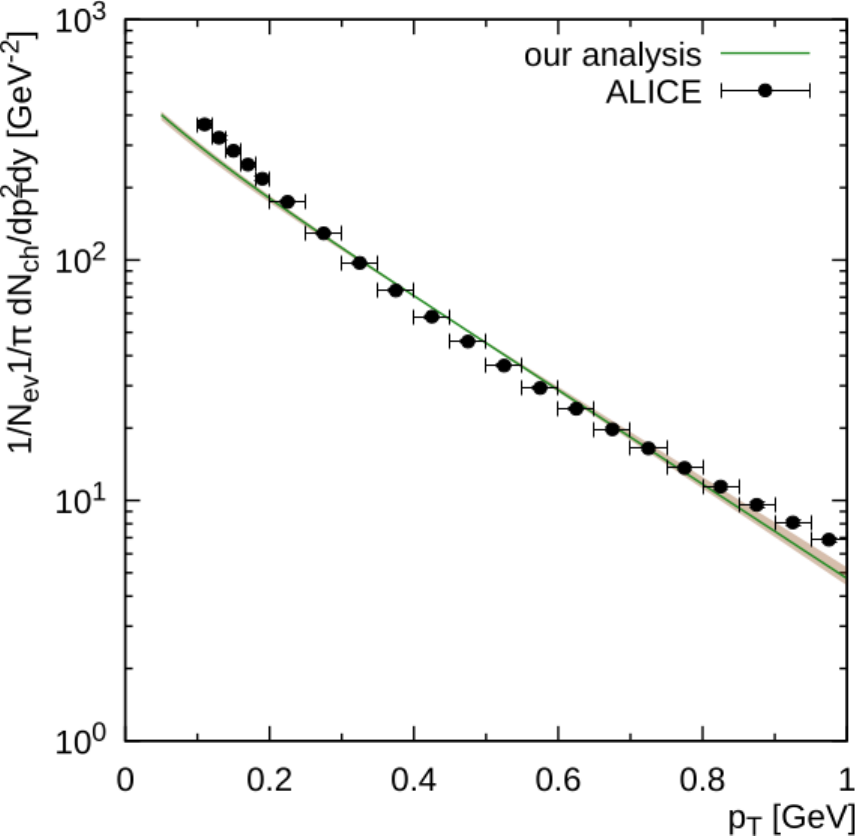}
        \includegraphics[width=5.0cm]{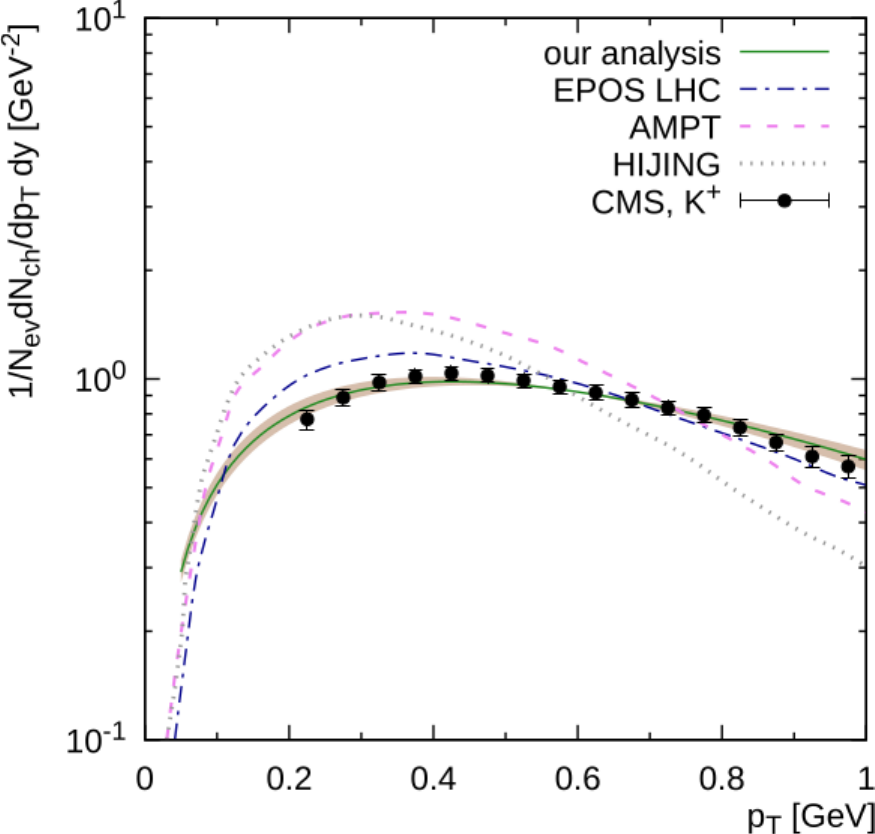}
        \includegraphics[width=5.0cm]{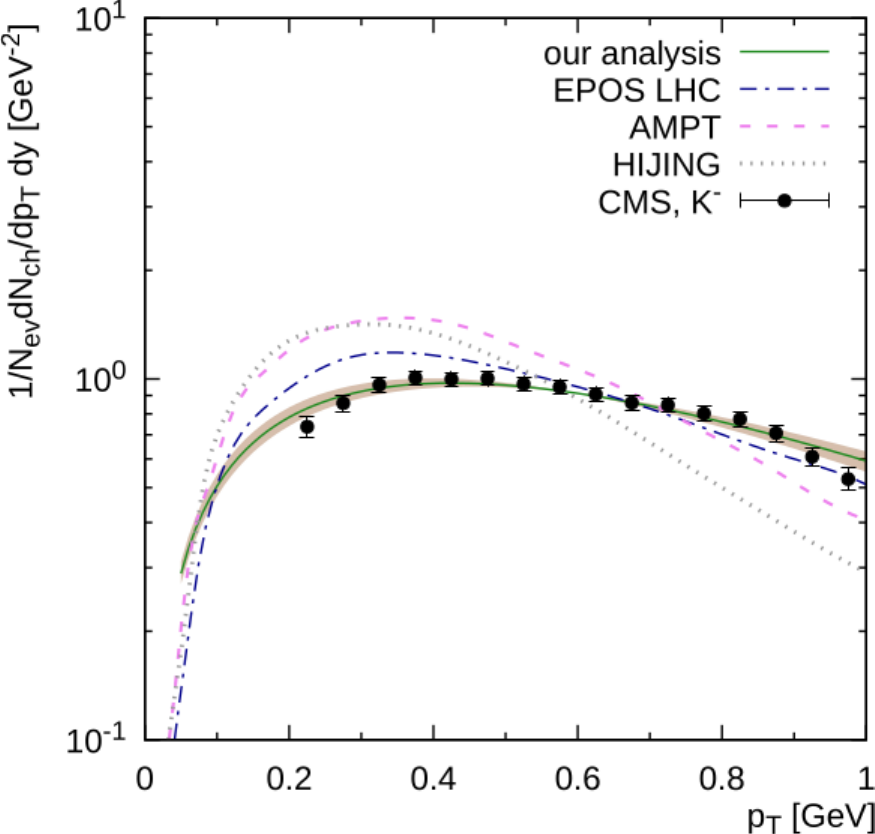}
        \includegraphics[width=5.0cm]{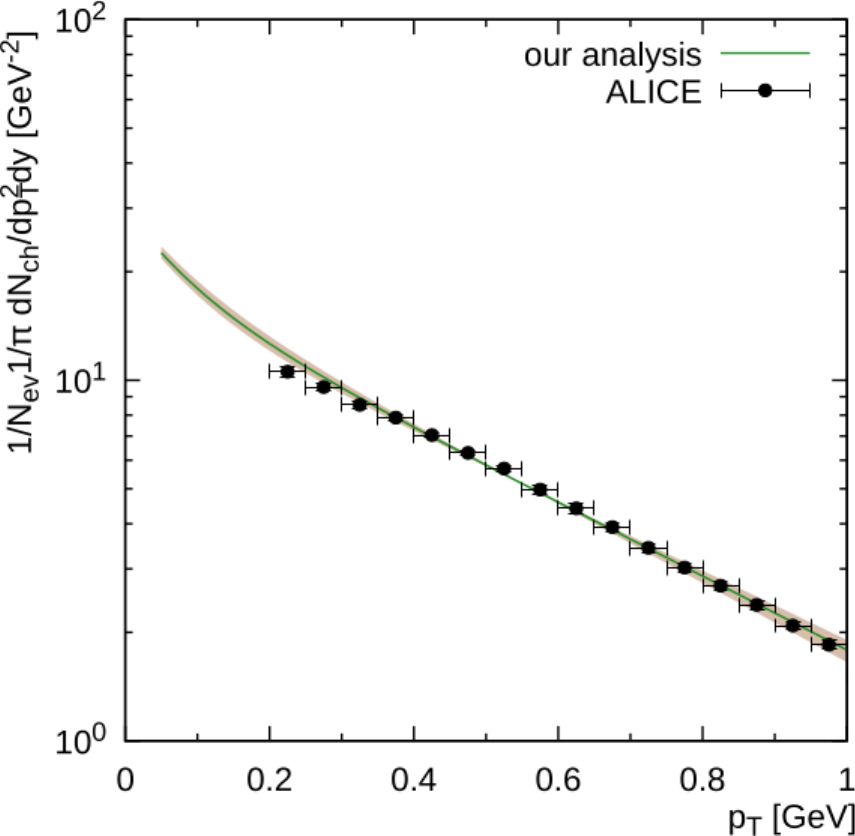}
        \caption{Pions (upper panels) and kaons (lower panels) multiplicities transverse momenta spectra calculated with nuclear LLM-2024 TMD compared with available LHC data
                for $p_T < 1$~GeV.
                The uncertainties of the fit procedure are shown as brown bands.
                The data of CMS~\cite{ChargedHadron5-CMS}, ATLAS~\cite{ChargedHadron5-ATLAS} and ALICE~\cite{ChargedHadron5-ALICE} were taken at $\sqrt{s}=$ 5.02~TeV for p-Pb collisions and summed over the collision centralities.}
        \label{fig2}
\end{figure}

In Fig.~\ref{fig2} we present our results for the  inclusive charged hadrons spectra
  measured in $p$-Pb collisions at $\sqrt{s} = 5.02$~TeV  at the central rapidities (in the center-of-mass frame) compared to CMS\cite{ChargedHadron5-CMS}, ATLAS\cite{ChargedHadron5-ATLAS} and ALICE\cite{ChargedHadron5-ALICE}. We sum the data over the centralities of the collisions
  and limit to $p_T < 1$~GeV.
The solid curves represent our predictions with brown bands being
 the uncertainties of the $B^a_h$ fits.
  At large $A=208$ the Fits~I and II almost coincide, so
 we took the Fit~I for definiteness. We find that best description of proton-lead data on $\pi$-mesons production can be
achieved with $B^g_\pi=9.1^{+0.2}_{-0.5}$, which is in agreement with our results obtained in $pp$ collisions, $B^g_\pi=8.7\pm 0.3$~\cite{LLM-2022,LLM-QFTHEP25}. A good agreement with the LHC data is achieved. A deviation from the data can be seen for $p_T\sim 1$~GeV for ALICE and ATLAS results, which may indicate the necessity to include
 QCD evolution effects.
 ALICE data also reveal some disagreement at very low $p_T < 0.2$~GeV,
 which may be connected with an underestimation of quark/diquark contributions.
 For $K$-mesons, we have obtained $B^g_K=4.8^{+0.4}_{-0.3}$. This value allows us to obtain a very good agreement with the data.
Additionally, we show the predictions from three Monte-Carlo event generators,
namely, \textsc{hijing}\cite{HIJING}, \textsc{ampt}\cite{AMPT} and \textsc{epos lhc}\cite{EPOS}. 
Our results for $\pi^\pm$ production appear to be very close to the \textsc{epos lhc} predictions
and the both agree well with the LHC data, better than agreement provided by the \textsc{ampt} and \textsc{hijing} tools.
For kaons we achieve a best description of the CMS data in comparison with the Monte-Carlo generators.
The results
show that the nuclear modification of the proposed TMD gluon density~\cite{nLLM-2024} does not contradict
the available collider data on $pA$ events.

In summary, the proposed LLM-2024 TMD, which can simultaneously describe LHC data on charged hadrons multiplicities at small $p_T$ and a number of HERA and LHC data ($F_2 (x,Q^2)$, $b$-jet production, Higgs production etc.) was extended to the nuclear case using the geometric scaling approach. The new nuclear LLM-2024 TMD PDF correctly describes shadowing at small $x$ and can be used to describe inclusive charged hadron production spectra at low transverse moments measured in $pA$ collisions at LHC. A better description of $K$-mesons production in $pA$ collisions compared with Monte-Carlo calculations can be obtained. The data also allow to extract FF parameters for $\pi$ and $K$-mesons. In future we plan to extend the obtained nuclear TMD to higher scales in order to make it possible to describe a wider range of LHC processes.

{\sl Acknowledgements.} We are deeply grateful to the organizers of the QCD at the Extremes conference for their dedication and efforts in fostering an inclusive and closely connected scientific community. We thank participants of Weekly-Online-Meetings, in particular S.P.~Baranov and H.~Jung for their
interest to this work, very important comments and remarks.
This research has been carried out at the expense of the Russian Science Foundation grant No.~25-22-00066, https://rscf.ru/en/project/25-22-00066/.



\providecommand{\href}[2]{#2}\begingroup\raggedright\endgroup

\end{document}